\let\ifarxiv=\iftrue     
\pdfoutput=1

\ifarxiv

\documentclass[12pt,a4paper]{article}
\usepackage[a4paper,text={450pt,650pt},centering]{geometry}

\fi

\ifarxiv\else

\documentclass[11pt,a4paper]{article}
\usepackage{mathptmx}
\usepackage[a4paper,text={130mm,198mm}]{geometry}

\fi

\usepackage{amsmath,amssymb}
\usepackage[bookmarks=true,hyperfigures=true]{hyperref}
\usepackage{graphicx}
\usepackage[nosort]{cite}
\usepackage[bulletsep]{collref}

\def\dss{ {\delta_{\sigma\sigma'}} }
\def\pdss{ {\partial_\sigma\delta_{\sigma\sigma'}} }
\def\uu{ {\underline{{\mathbf{1}}}} }
\def\dd{ {\underline{{\mathbf{2}}}} }
\def\ttr{ {\underline{{\mathbf{3}}}} }
\def\LL{{\cal L}}
\def\beq{\begin{equation}}
\def\eeq{\end{equation}}
\def\beqz{\begin{equation*}}
\def\eeqz{\end{equation*}}
\def\bea{\begin{eqnarray}}
\def\eea{\end{eqnarray}}
\def\nn{\nonumber}
\def\tr{ {\mbox{Tr} }}
\def\str{ {\mbox{Str} }}

\def\et{\qquad\mbox{and}\qquad}
\def\half{\frac{1}{2}}
\def\Ab{{\bar{A}}}
\def\pb{{\bar{\partial}}}
\let\oldbfseries=\bfseries
\let\oldmdseries=\mdseries
\let\oldnormalfont=\normalfont
\renewcommand{\bfseries}{\oldbfseries\boldmath}
\renewcommand{\mdseries}{\oldmdseries\unboldmath}
\renewcommand{\normalfont}{\oldnormalfont\unboldmath}

\allowdisplaybreaks[3]

\numberwithin{equation}{section}

\usepackage[font=small,labelfont=bf,width=0.85\textwidth]{caption}

\providecommand{\hypersetup}[1]{}

\hypersetup{plainpages=false}
\hypersetup{pdfpagemode=UseNone}
\hypersetup{bookmarksnumbered=true}
\hypersetup{pdfstartview=FitH}
\hypersetup{colorlinks=false}
\hypersetup{citebordercolor={.5 1 .5}}
\hypersetup{urlbordercolor={.5 1 1}}
\hypersetup{linkbordercolor={1 .7 .7}}

\providecommand{\arxivref}[2]{\href{http://arxiv.org/abs/#1}{#2}}
\providecommand{\doiref}[2]{\href{http://dx.doi.org/#1}{#2}}

\providecommand{\href}[2]{#2}
\providecommand{\arxivlink}[1]{\href{http://arxiv.org/abs/#1}{arxiv:#1}}

\begin{document}


\thispagestyle{empty}
\phantomsection
\addcontentsline{toc}{section}{Title}

\begin{flushright}\footnotesize%
\texttt{AEI-2010-132},
\texttt{\arxivlink{1012.3988}}\\
overview article: \texttt{\arxivlink{1012.3982}}%
\vspace{1em}%
\end{flushright}

\begingroup\parindent0pt
\begingroup\bfseries\ifarxiv\Large\else\LARGE\fi
\hypersetup{pdftitle={Review of AdS/CFT Integrability, Chapter II.3: Sigma Model, Gauge Fixing}}%
Review of AdS/CFT Integrability, Chapter II.3:\\
Sigma Model, Gauge Fixing\par\endgroup
\vspace{1.5em}
\begingroup\ifarxiv\scshape\else\large\fi%
\hypersetup{pdfauthor={Marc Magro}}%
Marc Magro
\par\endgroup
\vspace{1em}
\begingroup\itshape
Universit\'e de Lyon, Laboratoire de Physique, ENS Lyon et CNRS UMR 5672,\\
46 all\'ee d'Italie, 69364 Lyon CEDEX 07, France
\medskip

and

\medskip
Max-Planck-Institut f\"ur Gravitationsphysik\\%
Albert-Einstein-Institut\\%
Am M\"uhlenberg 1, 14476 Potsdam, Germany
\par\endgroup
\vspace{1em}
\begingroup\ttfamily
Marc.Magro@ens-lyon.fr
\par\endgroup
\vspace{1.0em}
\endgroup

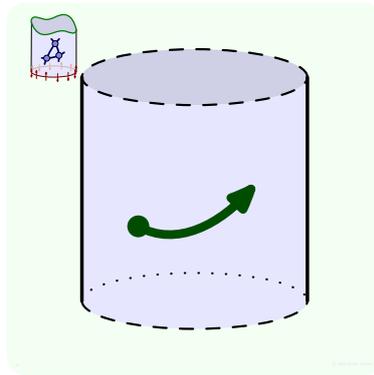
\begin{center}
\includegraphics[width=5cm]{TitleII3.mps}
\vspace{1.0em}
\end{center}

\paragraph{Abstract:}
This review is devoted to the classical integrability of the $\mbox{AdS}_5 \times \mbox{S}^5$ superstring theory. It starts with a reminder of the corresponding action as a coset model. The symmetries of this action are then reviewed. The classical integrability is then considered from the lagrangian and hamiltonian points of view. The second part of this review deals with the gauge fixing of this theory. Finally, some aspects of  the pure spinor formulation are also briefly reviewed.

\ifarxiv\else
\fi
\hypersetup{pdfsubject={MSC (2010): 81T30, 83E30, 81U15}}%

\ifarxiv\else
\fi
\hypersetup{pdfkeywords={AdS/CFT, integrable models}}%

\newpage


\section{Introduction}

The list of topics reviewed in this chapter is the following. First, we review the classical action of $AdS_5 \times S^5$ superstring theory as a supercoset model and its symmetries. The next topic concerns the integrability of that model and two important objects related to it, the Lax pair and the monodromy matrix. For all these aspects, a key role is played by a $\mathbb{Z}_4$ grading of the superalgebra $\mathfrak{psu}(2,2|4)$. The integrability property is then discussed from a Hamiltonian point of view. More precisely, it is recalled how to prove that an infinite number of conserved quantities are in involution. The first part of this chapter ends by recalling how factorized scattering theory is used in the quantum case. The second part of the review deals with gauge fixing, in particular with the so called uniform light-cone gauge, which is adapted to apply factorized scattering theory and to test the AdS/CFT conjecture. This chapter ends with some aspects related to the pure spinor formulation.

\paragraph{Note} The topics reviewed here are restricted on purpose. The main references related to these topics are indicated in the last section.

\section{Classical integrability}

\subsection{Action as a coset model and its symmetries}
\label{sec21}
\paragraph{Metsaev-Tseytlin Action}
The action is of the sigma-model type   on the coset superspace\footnote{More precisely, one needs to consider the universal cover as the physical space $AdS_5$ is a universal cover, see \cite{chapSuperconf}.}
\beq
G/H = PSU(2,2|4) \big/ \bigl[SO(4,1) \times SO(5)\bigr], \label{thecoset}
\eeq
together with  a Wess-Zumino term \cite{Metsaev:1998it}. This is therefore a generalization of the situation encountered in the flat case \cite{Henneaux:1984mh}. The bosonic part of the coset defined by (\ref{thecoset}) is $SO(4,2)/SO(4,1) \times SO(6) /SO(5)$ which corresponds to $AdS_5 \times S^5$. The Lie superalgebra ${\mathfrak su}(2,2|4)$ is a non-compact real form of $\mathfrak{sl}(4|4)$, which can itself be spanned by the $8 \times 8$ matrices   written in $4 \times 4$ blocks and whose supertrace (Str) vanishes. Here $\str \, M = \tr \, A - \tr \, D$ where $A$ and $D$ are the top and bottom diagonal $4 \times 4$ blocks of the matrix $M$. The superalgebra ${\mathfrak g} = {\mathfrak psu}(2,2|4)$ is then obtained by quotienting ${\mathfrak su}(2,2|4)$ over the ${\mathfrak u}(1)$ factor corresponding to the identity. In the following,  $\{t_A\}$  denotes a corresponding basis of  ${\mathfrak g}$, $\eta_{AB} = \str(t_At_B)$ and  $\eta^{AB}$ its inverse.

The coset (\ref{thecoset}) is associated with an automorphism $\Omega$ of order 4 of ${\mathfrak g}$. This means that ${\mathfrak g}$ admits a $\mathbb{Z}_4$ grading:
\begin{equation}
 {\mathfrak g} = {\mathfrak g}^{(0)} \oplus {\mathfrak g}^{(1)} \oplus {\mathfrak g}^{(2)} \oplus {\mathfrak g}^{(3)}
\end{equation}
with ${\mathfrak g}^{(0)} = {\mathfrak h} =  \mathfrak{so}(4,1) \oplus \mathfrak{so}(5)$ and $[ {\mathfrak g}^{(m)} ,  {\mathfrak g}^{(n)} ] \subset  {\mathfrak g}^{(p)}$ with $ p = m+n$ mod 4. The generators of  ${\mathfrak g}^{(0)}$ and ${\mathfrak g}^{(2)}$ are  even   while those of ${\mathfrak g}^{(1)}$ and ${\mathfrak g}^{(3)}$ are odd.  The supertrace is compatible with the $\mathbb{Z}_4$ grading, which means that $\str(M_m M_n) =0$ for $M_m \in \mathfrak{g}^{(m)}$, $M_n \in \mathfrak{g}^{(n)}$ and $m+n\neq 0$ mod $4$.

Let $(\sigma, \tau)$ be coordinates on the world-sheet and  $g(\sigma,\tau)$ a periodic function, $g(\sigma+\ell,\tau)=g(\sigma,\tau)$,  taking values in $G$.   The Lagrangian is written in terms of the left-invariant current $A_\alpha = - g^{-1} \partial_\alpha g$:
\beq
L= -\frac{\sqrt{\lambda}}{4 \pi}   \str\left[ \gamma^{\alpha\beta} A_\alpha^{(2)} A_\beta^{(2)} + \kappa \epsilon^{\alpha\beta} A_\alpha^{(1)} A_\beta^{(3)}\right]. \label{magro-action}
\eeq
Here,  $\epsilon^{\alpha\beta}$ is antisymmetric with $\epsilon^{\tau\sigma}=1$; $\gamma^{\alpha\beta}$ is the Weyl-invariant combination of the world-sheet metric with $\mbox{det} \, \gamma =-1$. For convenience, the coefficient in front of the Lagrangian has been written in terms of the t'Hooft coupling constant $\lambda$, with the AdS/CFT correspondence $\sqrt{\lambda} \leftrightarrow (R^2/\alpha')$, where $R$ is the common radius of $S^5$ and $AdS_5$ and $\alpha'$ the string slope.

The first term of the action corresponds simply to a non-linear sigma model on $AdS_5 \times S^5$. The second term is like a Wess-Zumino term which relies on the $\mathbb{Z}_4$ decomposition of ${\mathfrak g}$.  This comes from the property\footnote{Using form notations.} \cite{Berkovits:1999zq}
\beqz
2 \, \str ( A^{(2)} \wedge A^{(3)} \wedge A^{(3)} - A^{(2)} \wedge A^{(1)} \wedge A^{(1)} ) = \mbox{d} \, \str ( A^{(1)} \wedge A^{(3)} )
\eeqz
which shows that the l.h.s. is a closed and exact 3-form and explains the 2d expression of the Wess-Zumino term. The coefficient $\kappa$ in front of this Wess-Zumino term is in fact equal to $\pm 1$ in order to have $\kappa$-symmetry (see below).

\paragraph{Equations of motion and global $PSU(2,2|4)$ symmetry}

By varying the action with respect to $g$, one finds the following equation of motion:
\beq
\partial_\alpha S^\alpha - [A_\alpha, S^\alpha]=0 \label{eqmvt}
\eeq
where $S^\alpha =   \gamma^{\alpha\beta} A_\beta^{(2)} - \frac{1}{2}  \epsilon^{\alpha\beta} (A_\beta^{(1)} - A_\beta^{(3)})$. By definition, the current $A_\alpha$ is also a solution of the Maurer-Cartan equation
\beq
\partial_0 A_1 - \partial_1 A_0 -[A_0,A_1] =0.  \label{mcmardi}
\eeq

\medskip

$A_\alpha$ being the left-invariant current,  the action corresponding to (\ref{magro-action}) is invariant under the global  transformation $g(\sigma,\tau) \to \tilde{g} g(\sigma,\tau)$ with $\tilde{g} \in PSU(2,2|4)$.
 The equation of motion (\ref{eqmvt}) is identical to the equation of conservation of the Noether current $(\sqrt{\lambda}/2\pi) g S^\alpha g^{-1}$ associated with that symmetry. The corresponding Noether charge and its projection onto an  element $M \in \mathfrak{psu}(2,2|4)$ are respectively
\beq
 Q= \frac{\sqrt{\lambda}}{2\pi} \int_0^\ell d\sigma g S^0 g^{-1} \et Q_M = \str(Q M). \label{project}
\eeq

\paragraph{$SO(4,1)\times SO(5)$ gauge symmetry}

Under a local right multiplication $g(\sigma,\tau) \to g(\sigma,\tau) h(\sigma,\tau)$ with $h(\sigma,\tau) \in H$, the components $A_\alpha^{(1,2,3)}$ of the current transform as $A_\alpha^{(1,2,3)} \to h^{-1} A_\alpha^{(1,2,3)} h$. This shows that the action is invariant under these $SO(4,1)\times SO(5)$ gauge transformations.

\paragraph{Virasoro constraints and reparametrization}

Varying the action with respect to the metric gives the Virasoro constraints
\beq
\str(A_\alpha^{(2)} A_\beta^{(2)}) - \half \gamma_{\alpha\beta} \gamma^{\rho\sigma} \str(A_\rho^{(2)} A_\sigma^{(2)}) = 0 \label{virc2}
\eeq
which reflect the two-dimensional reparameterization invariance of the action.

\paragraph{$\kappa$-symmetry}

This symmetry is a key property of the Green-Schwarz formulation of superstring theories as it enables the reduction of the fermionic degrees of freedom to the physical ones. It acts on both the group element $g$ and the world-sheet metric $\gamma^{\alpha\beta}$. Its action on $g$ can be viewed as a particular local right multiplication  that depends on fermionic parameters \cite{McArthur:1999dy}. More precisely, at the infinitesimal level, it corresponds to $\delta g = g (\epsilon^{(1)} + \epsilon^{(3)})$ with\footnote{See \S 6.1 of \cite{Grigoriev:2007bu} or  \S 1.2.3 of \cite{Arutyunov:2009ga} for more details.}
\beqz
\epsilon^{(1)} = i A_{\alpha,+}^{(2)} \kappa^{(1)\alpha}_- + i\kappa^{(1)\alpha}_- A_{\alpha,+}^{(2)} \et  \epsilon^{(3)}  = i A_{\alpha,-}^{(2)} \kappa^{(3)\alpha}_+ + i\kappa^{(3)\alpha}_+ A_{\alpha,-}^{(2)}.
\eeqz
In these equations, $V_\pm^{\alpha} \equiv \frac{1}{2} (\gamma^{\alpha\beta} \mp  \epsilon^{\alpha\beta}) V_\beta$, $\kappa^{(1)}_+ =0$ and $\kappa^{(3)}_- =0$. The corresponding transformation of the metric can be written as:
\beqz
\delta \gamma^{\alpha\beta} = -\half \str\Bigl(W \bigl([i\kappa_{-}^{(1)\alpha}, A_-^{(1)\beta}] + [i\kappa^{(3)\alpha}_+ , A^{(3)\beta}_+]\bigr)\Bigr)
\eeqz
where $W$ is the diagonal matrix $(1,\cdots,1,-1,\cdots,-1)$.

\subsection{Lagrangian integrability}
\paragraph{Lax pair and monodromy} The  requirement for classical integrability is the existence of an infinite number of conserved quantities. This is ensured when the equations of motion are equivalent to a zero curvature equation
\begin{equation}
 \partial_\alpha L_\beta - \partial_\beta L_\alpha - [L_\alpha, L_\beta] = 0 \label{magro01}
\end{equation}
associated with  a Lax connection $L_\alpha(\sigma,\tau,z)$ depending on the dynamical fields and on a complex spectral parameter $z$.   Indeed, a consequence of this equation is that the monodromy
\begin{equation}
 T(\tau,z) = \stackrel{\longleftarrow}{\exp} \int_0^{\ell} d\sigma L_\sigma(\sigma,\tau,z) \label{defmon}
\end{equation}
satisfies the equation
\beqz
 \partial_\tau T(\tau,z) = [L_\tau(0,\tau,z),T(\tau,z)].
\eeqz

Therefore, its eigenvalues, which depend on the complex spectral parameter $z$, form an infinite set of  conserved quantities. Let us remark that for an integrable model  defined on the 2d plane rather than on the cylinder,  the time evolution of the corresponding monodromy  matrix obeys the equation
\beqz
\partial_\tau T(\tau,z) = L_\tau(+\infty,\tau,z)T(\tau,z) - T(\tau,z) L_\tau(-\infty,\tau,z).
\eeqz
However, to have configurations of finite energy, one has typically $L_\tau(\sigma,\tau,z)  \to 0$ when $\sigma \to \pm \infty$. If it is the case, then the whole monodromy is conserved.

A Lax connection is not unique and one can construct other Lax connections by making a formal gauge transformation\footnote{In the present case, $U \in PSU(2,2|4)$.}
\begin{equation}
L_\alpha  \to U L_\alpha U^{-1} + \partial_\alpha U U^{-1} \label{formalgauge}
\end{equation}
where $U(\sigma+\ell,\tau)=U(\sigma,\tau)$. The eigenvalues of $T(\tau,z)$ are invariant under such transformations.

The fact that $AdS_5 \times S^5$ superstring theory is integrable is not the sole peculiarity of  this theory.  It  originates in the existence of the associated $\mathbb{Z}_4$ grading  and is a generalization of the situation encountered in the bosonic case for a symmetric coset corresponding to a $\mathbb{Z}_2$ grading. To prove the existence of a Lax connection, one can start with an ansatz for $L_\alpha(z)$ generalizing the situation for the symmetric spaces\footnote{We use here form notations and $\ast$ designates the Hodge star on the worldsheet.},
\beqz
L(z) = a_1(z)  A^{(0)} + a_2(z)  A^{(2)} + a_3(z) \ast A^{(2)} + a_4(z) A^{(1)} +a_5(z) A^{(3)},
\eeqz
and determine the conditions on the coefficients $a_i(z)$ in order the flatness condition (\ref{magro01}) to reproduce the Maurer-Cartan equations (\ref{mcmardi}) and the equation of motion (\ref{eqmvt}). Proceeding like that, one can show that the quantity
\begin{equation}
L(z) = A^{(0)} + z^{-1} A^{(1)} + \frac{1}{2} (z^2 + z^{-2}) A^{(2)} + \frac{1}{2} (z^2 - z^{-2}) \ast A^{(2)}+ z A^{(3)}\label{magro02}
\end{equation}
is a Lax connection \cite{Bena:2003wd}.

\paragraph{$\kappa$-symmetry and integrability}

As previously mentioned, the theory is  invariant under $\kappa$-symmetry transformations only when the parameter $\kappa$ in front of the Wess-Zumino term  equals $\pm 1$. The existence of a Lax connection or, in other words, the integrability of the theory, is only valid for the  same values of $\kappa$. One rough way to understand this fact is that the corresponding bosonic coset model is integrable. This integrability property is thus extended to the full Green-Schwarz action, via the $\kappa$-symmetry, which relates bosons to fermions. It is also possible to prove that under a $\kappa$-symmetry transformation, and using the Virasoro constraints (\ref{virc2}), the Lax connection (\ref{magro02}) undergoes a formal gauge transformation (\ref{formalgauge}). This shows that the eigenvalues of the monodromy matrix are $\kappa$-symmetry invariant. Note that it is also clear that these eigenvalues are invariant under a $SO(4,1) \times SO(5)$ gauge transformation.

\paragraph{Local and non-local conserved charges}

The conserved charges are both local and non-local. Typically, they can be obtained by expansion around some particular value of the spectral parameter.  One can obtain for instance  a sequence of  local charges. Another possible sequence starts with the Noether charges (\ref{project})  and goes on with multi-local charges. This discussion is closely related to the study of the algebraic curve \cite{Beisert:2005bm}, which is associated with the eigenvalues of $T(z)$. It is also related to the construction of the Yangian charges. We refer to \cite{chapCurve}, \cite{chapYang} and more generally to \cite{Babebook}.

\subsection{Hamiltonian integrability}

\paragraph{Canonical analysis}

At the Hamiltonian level, a "conservative"  definition of integrability requires a further condition. There must be an infinite number of conserved quantities that are in involution, which means that their Poisson brackets (P.B.) vanish. For finite dimensional systems, this condition is necessary in order
to apply Liouville's theorem. The proof that such a property holds for string theory on $AdS_5 \times S^5$ is rather technical and therefore only intermediate steps will be reviewed here.

The first step is to do a canonical analysis by considering the current $A_\alpha$ as a dynamical variable rather than the group element $g$ itself. Due to this choice and to the gauge invariances of the action, there are constraints on the phase space. Applying  the Dirac procedure for constrained systems, one finds that  the theory can be described by the spatial component $A_\sigma(\sigma,\tau)$ of this current and its  conjugate momentum $\Pi(\sigma,\tau)$ with four types of constraints. First the Virasoro constraints. Then a bosonic constraint,  ${\cal C}^{(0)}$, associated with the $SO(4,1) \times SO(5)$ gauge invariance. Finally, two fermionic constraints $({\cal C}^{(1)},{\cal C}^{(3)})$. It is possible to extract from each of the fermionic constraints two constraints, $({\cal K}^{(1)},{\cal K}^{(3)})$, which are first-class\footnote{Which means that their P.B. with all the other constraints vanish on the constraint surface.} and generate the $\kappa$-symmetry transformations. However, as usual with $\kappa$-symmetry, it is not possible to separate covariantly  $({\cal C}^{(1)},{\cal C}^{(3)})$ into $({\cal K}^{(1)},{\cal K}^{(3)})$ and a complementary set of second-class constraints.

Rather than $\Pi$ itself, the interesting quantity is in fact $(\nabla_\sigma \Pi )$ where $\nabla_\sigma = \partial_\sigma - [A_\sigma,]$. In the case of the principal chiral model, this can be understood as $(\nabla_\sigma \Pi )$ coincides with the time component $A_\tau$ of the current. The result of this analysis is that the P.B. of $A_\sigma$ and $(\nabla_\sigma \Pi )$ take the same form as in the principal chiral model. The most convenient way to write these P.B. is to use tensorial notation and to define for any quantity $M \in \mathfrak{g}$, $M_\uu = M \otimes 1$ and $M_\dd = 1 \otimes M$. Then, we have\footnote{The time dependence is not indicated in the P.B. as they are equal-time P.B.},
 \begin{align*}
\{ A_{\sigma\uu}(\sigma), A_{\sigma\dd}(\sigma') \} &= 0,\\
\{ (\nabla_\sigma \Pi)_{\uu}(\sigma), A_{\sigma\dd}(\sigma') \} &= \bigl[C_{\uu\dd}, A_{\sigma\dd} \bigr] \dss - C_{\uu\dd} \pdss,\\
\{ (\nabla_\sigma \Pi)_{\uu}(\sigma) ,  (\nabla_{\sigma} \Pi)_{\dd}(\sigma') \} &=
 \bigl[C_{\uu\dd},   (\nabla_\sigma \Pi)_{\dd} \bigr] \dss.
\end{align*}
The quadratic Casimir is defined by:
\begin{align*}
C_{\uu\dd} = \eta^{AB} t_A \otimes t_B= C_{\uu\dd}^{(00)} + C_{\uu\dd}^{(13)} + C_{\uu\dd}^{(22)} + C_{\uu\dd}^{(31)},
\end{align*}
where in the last equality we have projected into the different gradings. The important characteristic of these P.B. is the presence of a non-ultra local term, proportional to $\delta'$.

\paragraph{Hamiltonian Lax Connection}

The next step is to mimic  the procedure recalled above for the Lagrangian analysis. One can start with a general expression for an Hamiltonian Lax connection as a  linear combination of $A_\sigma^{(i)}$ and $(\nabla_\sigma \Pi)^{(j)}$. However, this does not fix completely the Lax connection and    leads to many different possibilities, that differ from each other by terms proportional to the constraints. It is nevertheless possible to determine a unique linear combination that satisfies the two following  conditions.   Firstly, that the zero curvature condition holds on the whole phase space, which means even without using the constraints. Secondly, that the conserved quantities  $\str[T^n(z)]$ obtained from the monodromy matrix associated with this particular Lax connection  are first-class, or, in other words, gauge-invariant. It is possible to show that the corresponding  $L_\alpha^H(z)$  differs from the corresponding Lagrangian expression (\ref{magro02}) by terms proportional to the constraints.

\paragraph{Poisson brackets of $L_\sigma^H(z)$}

The goal is to compute the Poisson brackets of the monodromy matrix associated with $L_\sigma^H(z)$. This requires first to compute the Poisson brackets of two spatial Lax components. This computation is straightforward. However, organizing the result in a specific algebraic form is much more difficult.  Denoting ${\cal L}(\sigma,z) \equiv L^H_\sigma(\sigma,\tau;z)$, the result of this analysis is
\begin{multline}
\{ {\cal L}_\uu(\sigma,z_1), {\cal L}_\dd(\sigma',z_2)\} = [r^-_{\uu\dd}(z_1,z_2),
{\cal L}_\uu(\sigma,z_1)] \delta_{\sigma\sigma'} + [r^+_{\uu\dd}(z_1,z_2),
{\cal L}_\dd(\sigma,z_2)] \delta_{\sigma\sigma'} \\
 - \bigl(r^+_{\uu\dd}(z_1,z_2) - r^-_{\uu\dd}(z_1,z_2)\bigr)\pdss. \label{magro-mail}
\end{multline}
The matrices $r_{\uu\dd}^\pm$ have the following expression:
\beqz
 r_{\uu\dd}^-(z_1,z_2) = \frac{2\sum_{j=0}^3 z_1^j z_2^{4-j} C_{\uu\dd}^{(j \,\, 4-j)}}{\phi(z_2)(z_2^4 - z_1^4)}, \qquad
r_{\uu\dd}^+(z_1,z_2) = \frac{2\sum_{j=0}^3 z_1^{4-j} z_2^{j} C_{\uu\dd}^{(4-j \,\, j)}}{\phi(z_1)(z_2^4 - z_1^4)} \nonumber
\eeqz
with $\phi(z) = z (du/dz)$ where
\beq
u(z) = 2\frac{1+z^4}{1-z^4} \label{zukho}
\eeq
 is the Zhukovsky map. The form (\ref{magro-mail}) of the P.B. is exactly similar to the one appearing in  the principal chiral model \cite{Maillet:1985fn}, \cite{Maillet:1985ek}. It is again non ultra-local due to the presence of the $\delta'$ term. The Jacobi identity for the Poisson bracket (\ref{magro-mail}) is ensured by the following property
\beq
[r_{\uu\dd}^- , r_{\uu\ttr}^-] + [r_{\uu\dd}^- , r_{\dd\ttr}^-] + [r_{\ttr\dd}^- , r_{\uu\ttr}^-] =0 \label{extendedyb}
\eeq
satisfied by $r_{\uu\dd}^-$.

\paragraph{Algebraic interpretation and the Zhukovsky map}

As usual with integrable models, it is also possible and instructive to start the story from a purely algebraic point of view. In this framework, the approach corresponds to the so-called $R$-matrix one. This means to construct first the $r_{\uu\dd}^\pm$ matrices independently of the model considered i.e. without any reference to phase-space variables. The realization in terms of phase space variables is then achieved at the end {\em via} the matrix ${\cal L}(\sigma,z)$.

Starting from $\mathfrak{g}=\mathfrak{psu}(2,2|4)$, one considers its loop algebra\footnote{More precisely, it is necessary to consider its twisted loop algebra $\mathfrak{Lg}^\Omega$, where the twist is induced by the $\mathbb{Z}_4$-grading $\Omega$ of $\mathfrak{g}$.}  $\mathfrak{Lg} = \mathfrak{g}[[z,z^{-1}]]$. Any $X(z) \in \mathfrak{Lg}$ can be decomposed into its pole part, $\pi_-(X)$ and its regular part $\pi_+(X)$. This splitting of $\mathfrak{Lg}$ enables one to define a $R$-matrix on $\mbox{End} \,\, \mathfrak{Lg}$. It is simply given by $R = \pi_+ - \pi_-$ and satisfies the modified classical Yang-Baxter equation:
\beq
[R(X), R(Y)] -R([R(X),Y] +[X, R(Y)]) = -[X,Y]. \label{modyb}
\eeq
Let $(\, . \,, \,. \,)$  be an inner-product on $\mathfrak{Lg}$. This inner product has a natural extension on $\mathfrak{Lg} \otimes \mathfrak{Lg}$. One can then associate to any operator $\mathfrak{O}  \in \mbox{End} \,\, \mathfrak{Lg}$ its kernel $\mathfrak{O}_{\uu\dd} \in \mathfrak{Lg} \otimes \mathfrak{Lg}$ through the relation
\beqz
\forall X, Y \in \mathfrak{Lg}, \qquad (\mathfrak{O}(X), Y) = (\mathfrak{O}_{\uu\dd}, Y \otimes X).
\eeqz
An important property is that the kernel of $\mathfrak{O}^\ast$ is simply\footnote{ $\mathfrak{O}_{\dd\uu} = P(\mathfrak{O}_{\uu\dd})$ with $P(A \otimes B) = B \otimes A$.} $\mathfrak{O}_{\dd\uu}$. The eq.(\ref{modyb}) can then be rewritten successively as
\begin{align}
 [R_{\uu\dd}, R_{\uu\ttr}] + [R_{\uu\dd}, R_{\dd\ttr}] +[R_{\ttr\dd}, R_{\uu\ttr}] &= - \widehat{\omega}_{\uu\dd\ttr}, \nn\\
[R_{\uu\dd}, R_{\uu\ttr}] + [R_{\uu\dd}, R_{\dd\ttr}]  -[R_{\uu\ttr}, R^\ast_{\dd\ttr}] &= - \widehat{\omega}_{\uu\dd\ttr}. \label{couldbeyb}
\end{align}
For simplicity the expression of $\widehat{\omega}$ is not reproduced here (see \cite{Vicedo:2010qd}).

The key point is the following: if we take the inner product
\beqz
(X,Y)_u = \oint \frac{du}{2 \pi i} \str\left(X(z) Y(z)\right)
\eeqz
with $u(z)$ given by (\ref{zukho}), then $R^\ast \neq -R$. This means that the eq.(\ref{couldbeyb}) does not correspond to the classical Yang-Baxter equation but to\footnote{The r.h.s. $-\widehat{\omega}$ is a contact term proportional to $\delta(z_1-z_2) \delta(z_2-z_3)$ and is absent in  eq.(\ref{extendedyb}).} the eq.(\ref{extendedyb}) with $R_{\uu\dd} = r_{\uu\dd}^-$ and $R^\ast_{\uu\dd} = - r_{\uu\dd}^+$. Therefore, the integrable structure of the $AdS_5 \times S^5$ superstring fits precisely into the general $R$-matrix approach. The specifics of this model are encoded in its hamiltonian Lax matrix which can be formally written as
\beqz
{\cal L}(\sigma,z) = 4 \phi(z)^{-1} \sum_{k=1}^\infty z^k \Bigl( k A_1^{(k)} + 2 (\nabla_\sigma \Pi)^{(k)} \Bigr).
\eeqz

\paragraph{Involution of the conserved quantities}

The last step, which is the computation of the P.B. of the monodromy matrix (\ref{defmon}) from the  result (\ref{magro-mail}) is delicate. Indeed, the non ultra-local term in  (\ref{magro-mail}) leads  to ambiguities  for the P.B. of the monodromy. The way to proceed is the following. Consider the transition matrices
\beqz
T(\sigma_1,\sigma_2,\tau,z) = \stackrel{\longleftarrow}{\exp}  \int_{\sigma_2}^{\sigma_1} d\sigma \LL(\sigma,z).
\eeqz
The P.B. of two transition matrices with all different points are well defined. However, there are ambiguities whenever two points coincide. A simple argument to understand this property is the following. To compute the P.B. of $T(\sigma_1,\sigma_2,\tau,z)$ with $T(\sigma'_1,\sigma'_2,\tau,z')$, we have, schematically, to twice integrate a $\delta'$ term multiplied by some smooth function. Omitting for brevity the function, we have to evaluate
\beqz
\int_{\sigma_2}^{\sigma_1} d\sigma \int_{\sigma'_2}^{\sigma'_1} d\sigma' \pdss  = \chi(\sigma_1 ; [\sigma'_2,\sigma'_1]) - \chi(\sigma_2;[\sigma'_2,\sigma'_1]),
\eeqz
where $\chi(\sigma;[\sigma',\sigma''])$ is the characteristic function of the interval $[\sigma',\sigma'']$. But this function is undefined when two points coincide. Therefore the P.B. of the monodromy matrices $T(\tau,z) =T(\ell,0,\tau,z)$   and $T(\tau,z')$ is not well defined.  However, it has been proved in \cite{Maillet:1985ek} that one can give a meaning to the limit of coinciding points if one imposes that the P.B. of the monodromy matrix satisfies the antisymmetry and the derivation  rules. This leads to a regularization which consists in  point splitting and in applying a symmetric limit procedure. This regularization  is equivalent to taking $\theta(0)=1/2$ where $\theta$ is the Heaviside function. This procedure leads to the following result for the P.B. of the monodromy matrix:
\beq
\{ T_\uu , T_\dd \} = \half [r^+_{\uu\dd} +r^-_{\uu\dd}  , T_\uu T_\dd] + \half T_\uu (r^+_{\uu\dd} -r^-_{\uu\dd}) T_\dd - \half T_\dd (r^+_{\uu\dd} -r^-_{\uu\dd}) T_\uu, \label{magrortst}
\eeq
where  $T_\uu=T(\tau,z_1) \otimes \text{Id}$ and $T_\dd = \text{Id} \otimes T(\tau,z_2)$. This P.B. is called the classical exchange algebra. Taking the supertrace on both spaces $\uu$ and $\dd$ , one finds that  the conserved quantities $\str[T^n(z_1)]$ and $\str[T^m(z_2)]$ are in involution. Let us however insist that contrary to what happens for the monodromy,  the P.B. of $\str[T^n(z_1)]$ and $\str[T^m(z_2)]$ has no ambiguity. In other words, its vanishing is independent of the choice of regularization.

\paragraph{What is the quantum exchange algebra ?}

The quantum analogue of the exchange algebra (\ref{magrortst}) is not known. This is in fact a long-standing problem for non ultra-local integrable models. The reason is that the P.B. (\ref{magrortst}) does not satisfy completely the Jacobi identity. This means that the P.B. $\{T_\uu,\{T_\dd, \{ \cdots,\cdots\}, T_{\underline{{\mathbf{n}}}}\}\}$ with $n$ occurrences of $T$ must be separately defined for each $n$. This is clearly an obstruction for the  determination of the quantum exchange algebra.

There are however integrable models for which the quantum exchange algebra is known. The simplest ones are of course ultra-local models. In that case  $r^+_{\uu\dd} = r^-_{\uu\dd}$ is an antisymmetric $r$-matrix, solution of the classical Yang-Baxter equation
\beq
[r_{\uu\dd} , r_{\uu\ttr}] + [r_{\uu\dd}, r_{\dd\ttr}] + [r_{\uu\ttr},r_{\dd\ttr}]=0. \label{ybeq}
\eeq
The quantum exchange algebra is then simply
$R_{\uu\dd} T_\uu T_\dd =  T_\dd T_\uu R_{\uu\dd}$  with $R_{\uu\dd} = 1 + \hbar r_{\uu\dd} + \cdots$ a solution of the quantum Yang-Baxter equation $R_{\uu\dd} R_{\uu\ttr}R_{\dd\ttr} = R_{\dd\ttr} R_{\uu\ttr} R_{\uu\dd}$. The interest of such a relation is that one can discretize the model and apply Bethe Ansatz techniques.

Another possibility, this time for some specific non ultra-local models, is when  the matrices $r^\pm_{\uu\dd}$ are such that $r_{\uu\dd} = (1/2) (r^+_{\uu\dd} +r^-_{\uu\dd})$ satisfies the classical Yang-Baxter equation (\ref{ybeq}). Denoting $s_{\uu\dd} = (1/2) (r^+_{\uu\dd} - r^-_{\uu\dd})$, the quantum analogue  of (\ref{magrortst}) for these models is \cite{Freidel:1991jx}, \cite{Freidel:1991jv}:
\beqz
R_{\uu\dd} T_\uu S_{\uu\dd} T_\dd = T_\dd S_{\uu\dd} T_\uu R_{\uu\dd}.
\eeqz
with $S_{\uu\dd} = 1 + \hbar s_{\uu\dd} + \cdots$ and similarly for $R$ and $r$. However, for both the principal chiral model and the superstring on $AdS_5 \times S^5$, the corresponding  matrix $r_{\uu\dd}$  does not satisfy the classical Yang-Baxter equation.

For  $AdS_5 \times S^5$ superstring theory, the only available results so far consist in the approach developed in  \cite{Mikhailov:2007eg} within the pure spinor formulation (see section \ref{secps}) and the subsequent conjecture made there. Some interesting results have however been obtained very recently in \cite{Benichou:2010ts} for conformal models on supergroups.

At this point, the results obtained in \cite{Dorey:2006mx}, \cite{Vicedo:2008jy} for the bosonic subsector $\mathbb{R} \times S^3$ of the full theory have to be mentioned. For this subsector, one has a similar P.B. as in eq.(\ref{magrortst}). However, on the space of finite-gap solutions, it is possible to show that some variables form a set of action-angle variables if one computes their P.B. from the expression (\ref{magrortst}). Such a result is interesting because it confirms the correctness of the expression for the action variables obtained from the algebraic curve.

\subsection{Quantum integrability and factorized scattering theory}

In order to compute the spectrum at the quantum level, one has therefore to follow another road. The idea is then to apply the methods of factorized scattering theory \cite{Zamolodchikov:1978xm}. The prerequisites are the following. As usual for quantization within the Green-Schwarz formulation, the  first step is to go to a light-cone gauge\footnote{Another possibility developed in \cite{Grigoriev:2007bu}, \cite{Mikhailov:2007xr} and subsequent articles for the full $AdS_5 \times S^5$ theory is to go to a conformal gauge and to make a Pohlmeyer reduction. This has the advantage of keeping manifest the 2d Lorentz invariance.}. In such a gauge, the theory has a massive spectrum. The idea is then to study first the decompactification limit by considering the theory on a plane instead of a cylinder. Since the theory has a massive spectrum, it makes sense to talk about a world-sheet $S$-matrix in that limit. Note however that the light-cone gauge action is not Lorentz invariant and therefore some properties must be adapted and extended to the case at hand.  The key hypothesis is to suppose that the theory remains integrable at the quantum level. This assumption means that the $n \to n$ $S$-matrix factorizes into a product of $2 \to 2$  $S$-matrices. Let us insist here that  it is in fact not necessary to have an infinite number of conserved quantities (see \cite{Dorey:1996gd} for a review).  The next step is then to determine the dispersion relation and the two-body $S$-matrix from the symmetries\footnote{More precisely the "off-shell" symmetries, see below.}  of the light-cone gauged action in the decompactification limit. Thus, an important question related to that program is to determine these symmetries. Once all these steps are completed, finite size effects can be considered. Here we review the first steps of this  procedure.

\section{Gauge Fixing}
\subsection{Motivation and choice of gauges}

In this review, we will mainly focus on the light-cone gauge that is most adapted to the program detailed above with the further requirement that it is suited for the comparison between the energy of string states and the conformal dimension of the dual $\mathcal{N}=4$ Yang-Mills operators.

Let us begin by recalling a few things about light-cone gauges. Consider first the purely bosonic case. In flat space, light-cone gauge fixing is realized in two steps. The first one consists in going to the conformal gauge $\gamma^{\alpha\beta} = \eta^{\alpha\beta}$. The second one is to fix the residual conformal diffeomorphism  symmetry by imposing $x_+(\sigma,\tau)=\tau$. Another way to implement these gauge fixing conditions is to use  the first-order formulation \cite{Goddard:1973qh} and to impose $x_+ = \tau$ and to fix $p_+$, the momentum conjugate of $x_-$, to a constant. If these two ways to proceed are equivalent in flat space, this is no more the case for a curved space. Furthermore, it is impossible to apply the first procedure in the case of $AdS_5 \times S^5$, in particular because its null Killing vectors are not covariantly constant \cite{Horowitz:1990sr}. Therefore, the bosonic light-cone gauge conditions are imposed within the first-order formulation.

As recalled in \cite{chapSpinning}, there are two inequivalent sets of null geodesics in $AdS_5 \times S^5$: for the first set, the geodesic stays  entirely in $AdS_5$, for the second one it wraps a big circle of $S^5$. These two possibilities correspond to two types of light-cone gauges. In the case of superstrings,  the $\kappa$-symmetry invariance must be also fixed and this leads again to different possibilities. Using the Poincar\'e coordinates patch, and viewing $\mathfrak{psu}(2,2|4)$ as  the four-dimensional $\mathcal{N}=4$ super-conformal algebra, one possibility to fix $\kappa$-symmetry is to set the fermions associated with the 16 superboost generators to zero. This gauge is called the S-gauge. It has been used in particular in \cite{Beisert:2008iq} for the study of  the 2d duality of $AdS_5 \times S^5$ related to the dual superconformal symmetry of scattering amplitudes in $\mathcal{N}=4$  super-Yang-Mills theory \cite{Drummond:2006rz}, \cite{Drummond:2008vq} (see \cite{chapDual}). Another possibility is to put to zero half of these fermions and half of the fermions associated with the supersymmetry generators. Combined to the AdS light-cone gauge, this leads to the action in \cite{Metsaev:2000yf} which is at most quartic in the fermions.

\subsection{Uniform light-cone gauge}

The gauge we will review leads to much more complicated action than the AdS light-cone gauge but is well suited for the AdS/CFT correspondence. Indeed, to test this conjecture, one needs to compare the space-time energy $E$ of a string state with the conformal dimension of the dual operator. One feature of the uniform light-cone gauge is precisely that the corresponding world-sheet Hamiltonian is simply related to $E$.  We only review here the main steps for the bosonic string on $AdS_5 \times S^5$   and refer to the literature for the complete treatment and for some subtleties omitted here.

\paragraph{Bosonic case}
(i) Consider first the metric in global coordinates
\beqz
ds^2 = R^2 \Biggl[- \Bigl( \frac{1+z^2/4}{1-z^2/4} \Bigr)^2 dt^2 + \frac{dz_i^2}{(1-z^2/4)^2} + \Bigl( \frac{1-y^2/4}{1+y^2/4} \Bigr)^2 d\phi^2 + \frac{dy_i^2}{(1+y^2/4)^2} \Biggr]
\eeqz
with $i=1,\cdots,4$ and where $(t,z_i)$ describe $AdS_5$, with $t$ the global time of $AdS_5$, while $(\phi,y_i)$ describe $S^5$, with $\phi$ an angle parameterizing the equator of $S^5$. The conserved charges associated with shifts in $t$ and $\phi$ are respectively the space-time energy $E = - \int_0^\ell d\sigma p_t $ and the angular momentum
$ J = \int_0^\ell d\sigma p_\phi$ where $p_t$ and $p_\phi$ are the  conjugate momenta respectively of $t$ and $\phi$. Define then $x_- = \phi -t$ and $x_+ = (1/2)(\phi +t)$, such that  $p_- = (p_\phi +p_t)$ and $p_+ = (1/2)(p_\phi - p_t)$. The corresponding conserved charges associated with these densities are
\beq
P_- = J-E \et P_+ = (1/2) (J+E). \label{rela1}
\eeq

(ii) The light-cone gauge conditions are
\beqz
x_+ = \tau \et p_+ = 1.
\eeqz
As a consequence of the last condition, the charge $P_+$ is identical to $\ell$.

(iii) The next step is to solve the Virasoro constraints. One of them gives $p_-$ in terms of a square root of the transverse coordinates\footnote{Formed by $z_i$, $y_i$ and their conjugate momenta.} $(x_M,p_M)$ while the other constraint together with the periodicity of the fields imply the following result for the world-sheet momentum $p_{WS}$ of the string:
\beqz
p_{WS} = - \int_0^\ell d\sigma p_M {x'}^M =  \int_0^\ell d\sigma x'_- = 0.
\eeqz
This condition is called the level-matching condition. In the dual picture, it corresponds to the vanishing of the  total momentum of multi-magnon configurations.

(iv) The Virasoro conditions being solved, the gauge-fixed Lagrangian is
\beqz
p_M \dot{x}^M + p_+ \dot{x}_- + p_-.
\eeqz
Since $p_+=1$, the second term in the r.h.s. is a total derivative, which means that the light-cone gauged action is of the form $\int (p_M \dot{x}^M - h)$ with the light-cone Hamiltonian density $h = - p_-(x_M,x'_M,p_M)$. Together with the relation (\ref{rela1}), this means that the light-cone Hamiltonian $H$ is identical to
\beq
H = - P_- = E-J. \label{lcham}
\eeq
This is the relation announced above between the space-time energy $E$, the light-cone Hamiltonian $H$ and the angular momentum $J$.

(v) The way to deal with the level-matching condition is to impose it on the states. In the dual picture,  this means for instance that double-magnon excitations can be considered. However, to correspond to a physical state, the two magnons should have opposite momenta. When the level-matching condition is imposed (respectively relaxed), one refers to the on-(off-)shell theory.

\paragraph{Full theory}

This short reminder does not reflect at all the difficulty when fermions are included ! In particular, some of the steps that need to be completed include  choosing an adequate coset representative (which is such that all the fermions are neutral under the isometries generated by shifts of $t$ and $\phi$), fixing the $\kappa$-symmetry gauge invariance and developing the first-order formulation for the complete Metsaev - Tseytlin action.

\paragraph{Decompactification limit}

As discussed above, in order to make use of the factorized scattering theory, the first step is to consider the decompactification limit, which means to go from the cylinder to the plane. As $\ell$ corresponds to $P_+$ this limit is obtained by letting $P_+ \to \infty$ while  keeping $\lambda$ fixed. Since the energies of the states are finite, the relations (\ref{rela1}) imply that $J$ goes to infinity in this limit.

\paragraph{Symmetry}

Let us first consider the case $P_+$ finite. It is clear from the results (\ref{lcham}) and  (\ref{rela1}) that the light-cone Hamiltonian and $P_+$ correspond  to  particular charges of the form (\ref{project}). More precisely, we have
\beqz
H =- \frac{i}{2} Q_{\Sigma_+} \et P_+ = \frac{i}{4}   Q_{\Sigma_-}
\eeqz
for some $\Sigma_\pm \in \mathfrak{psu}(2,2|4)$. As $x_+ =\tau$, all the charges $Q_M$ that are independent of $x_+$ and commute with $H$ are conserved. However, we have a general result
\beq
\{ H, Q_M \} = - \frac{i}{2} \{ Q_{ \Sigma_+}, Q_M\} =   - \frac{i}{2} Q_{[ \Sigma_+, M]}. \label{21m}
\eeq
Therefore, all the  elements $M \in \mathfrak{psu}(2,2|4)$ that commute with $\Sigma_+$ give conserved charges. It can be shown that these elements correspond to
\beqz
 \mathfrak{psu}(2|2) \oplus \mathfrak{psu}(2|2) \oplus \Sigma_+ \oplus \Sigma_-,
\eeqz
the two last elements being associated with $H$ and $P_+$.

We need now first to go to the decompactification limit and then off-shell. In the decompactification limit, we are left a priori with $\mathfrak{psu}(2|2) \oplus \mathfrak{psu}(2|2) $ together with a central charge that corresponds to the Hamiltonian. However, this is  not the final answer for the off-shell theory. The reason is that for odd elements $M_1$ and $M_2$, central charges may  appear in the Poisson bracket $\{Q_{M_1}, Q_{M_2} \} $. This means that this P.B. is only equal to $Q_{[M_1,M_2]}$ up to some central charges. An explicit computation enables one to determine these central charges and shows that the  symmetry is $\mathfrak{psu}(2|2) \oplus \mathfrak{psu}(2|2)$  extended by three central charges $H$, $C$ and $C^\dagger$ with
\beqz
C= \frac{i \sqrt{\lambda}}{4 \pi}   (e^{i p_{ws}} -1).
\eeqz
As it should be, $C$ vanishes when $p_{ws}=0$. The determination of the off-shell symmetry algebra is the starting point needed to apply factorized scattering theory.

\section{Pure spinor formulation}
\label{secps}
In this section, we mention some results that have been obtained for the pure spinor (P.S.) formulation and that are directly related to the aspects treated in this review for the Green-Schwarz (G.S.) formulation.

The Lagrangian can be written as\footnote{Taking the supertrace is understood.} \cite{Berkovits:2000fe}
\beqz
L= {1 \over 2} A^{(2)} \Ab^{(2)} + {1 \over 4} A^{(1)} \Ab^{(3)} + {3 \over 4} A^{(3)} \Ab^{(1)} + w \pb \lambda + \bar{w} \partial \bar{\lambda} - N \Ab^{(0)} - \bar{N} A^{(0)} - N \bar{N}.
\eeqz
It is written in conformal gauge. Here, $A = -g^{-1} \partial g$ with $\partial = \partial_0 + \partial_1$ while $\Ab = - g^{-1} \pb g$ with $\pb = \partial_0 - \partial_1$. The fields $\lambda$ and $\bar{\lambda}$ are bosonic ghosts taking values in ${\mathfrak g}^{(1)}$ and ${\mathfrak g}^{(3)}$ respectively.  They satisfy the pure spinor conditions:
\beqz
 [\lambda,\lambda]_+ =0 \qquad \mbox{and}\qquad [\bar{\lambda},\bar{\lambda}]_+=0.
 \eeqz
$w$ and $\bar{w}$, are the conjugate momenta respectively of $\lambda$ and $\bar{\lambda}$ and take values respectively in  ${\mathfrak g}^{(3)}$ and ${\mathfrak g}^{(1)}$. Finally, $N$ and $\bar{N}$ are the pure spinor currents defined by:
\beqz
N = - [w,\lambda]_+ = -w \lambda -\lambda w \et \bar{N} = - [\bar{w},\bar{\lambda}]_+= -\bar{w} \bar{\lambda} - \bar{\lambda} \bar{w}.
\eeqz
They take values in ${\mathfrak g}^{(0)}$. There are a $SO(4,1) \times SO(5)$ gauge  invariance and a global $PSU(2,2|4)$ invariance.  However, $\kappa$-symmetry is not present but there is an invariance under a  BRST symmetry $Q = \int \str(dz \lambda A_3 + d\bar{z}\bar{\lambda} \bar{A}_1)$.

The equations of motion can again be rewritten as a zero curvature equation $ \bar{\partial} {\cal L} - \partial\bar{ {\cal L}} -[\bar{{\cal L}}, {\cal L} ]=0$ for the Lax connection \cite{Vallilo:2003nx}
\begin{align*}
 {\cal L}(z)  &= \bigl(A^{(0)} +N -z^4 N\bigr) + zA^{(1)} + z^2 A^{(2)} + z^3 A^{(3)},\\
 \bar{{\cal L}}(z)  &= \bigl(\Ab^{(0)} +\bar{N} -z^{-4} \bar{N}\bigr) +  z^{-3} \Ab^{(1)} +  z^{-2} \Ab^{(2)} + z^{-1}\Ab^{(3)},
 \end{align*}
which means that the theory is classically integrable.  In  the G.S. formulation, the eigenvalues of the  monodromy matrix are $\kappa$-symmetry invariant. The corresponding statement in the P.S. formulation is that they are BRST invariant. When putting the ghosts to zero, the Lagrangian Lax pair is different from the one in (\ref{magro02}). However, it is possible again to determine an Hamiltonian Lax connection for the P.S. formulation. This connection agrees with the Hamiltonian one of the G.S. formulation up to terms proportional to the ghosts. As a consequence, the P.S. classical exchange algebra is the same as in the G.S. formulation and this property remains true when the contribution of the ghosts to the P.B. is included. This classical exchange algebra has been first obtained in\footnote{There is a subtlety in the actual comparison with the result  (\ref{magro-mail}) due to the fact that the observables considered in \cite{Mikhailov:2007eg}   are gauge-invariant.} \cite{Mikhailov:2007eg}.

As this review focuses on the classical case, we just indicate briefly some results related to the quantum case and which are directly relevant to the framework of this review. Contrary to the G.S. formulation where going to a light-cone gauge breaks the global $PSU(2,2|4)$ symmetry and the conformal invariance, the quantization of the P.S. action is done within the framework of a 2d conformal field theory with an unbroken $PSU(2,2|4)$ invariance.  It has been proved that at the quantum level this theory is conformally and BRST invariant \cite{Berkovits:2004xu}, \cite{Vallilo:2002mh}. Furthermore, the classically conserved non-local currents can be made BRST invariant at the quantum level \cite{Berkovits:2004jw}. The one-loop corrections to the tree level OPE \cite{Mikhailov:2007mr} \cite{Puletti:2006vb} \cite{Bianchi:2006im}   of the left-invariant currents have been studied in \cite{Bedoya:2010av} and it has been explicitly demonstrated in \cite{Mikhailov:2007mr} that the monodromy matrix  is not renormalized at one loop.

\section{References}

First of all, for many aspects treated in this chapter, the reader is referred to the extended pedagogical review \cite{Arutyunov:2009ga}. The reference \cite{Zarembo:2010sg} presents a systematic discussion of other string backgrounds that share the properties reviewed in \S\ref{sec21}. A general reference on integrable models is the book \cite{Babebook}. The classical exchange algebra was first obtained within the pure spinor formulation in \cite{Mikhailov:2007eg}. It was rederived within that formulation and within the Green-Schwarz formulation in
 \cite{Magro:2008dv}. The analysis to fix the Hamiltonian Lax connection has been presented in \cite{Vicedo:2009sn}. The algebraic origin and interpretation of the Hamiltonian Lax connection and of the $r_{\uu\dd}^\pm$ matrices have been put forward in \cite{Vicedo:2010qd}. General references about the $R$-matrix approach can be found in the bibliography of the latter. For earlier attempts to compute the classical exchange algebra, see \cite{Das:2004hy},  \cite{Mikhailov:2006uc}, \cite{Kluson:2007ua}, and \cite{Das:2005hp}, \cite{Aoyama:2007tz} in AdS light-cone gauge.  For the problem of non ultra-local terms we recommend  the thesis \cite{Vicedo:2008jk}. For the AdS light-cone gauge, we refer to the proceedings \cite{Tseytlin:2000na}, to  the original references \cite{Metsaev:2000yu}, \cite{Metsaev:2000yf} and to \cite{Alday:2005gi} for the integrability of the theory in that gauge. For the uniform light-cone gauge, the references for the topics reviewed here are \cite{Arutyunov:2004yx}, \cite{Arutyunov:2005hd}, \cite{Arutyunov:2006gs} and more specifically \cite{Frolov:2006cc}, \cite{Arutyunov:2006ak} and, once again, the review \cite{Arutyunov:2009ga}.  Further references are indicated in \cite{chapQString}. Finally, the references \cite{Adam:2007ws} and \cite{Puletti:2010ge} contain a pedagogical introduction to the P.S. formulation of $AdS_5 \times S^5$ superstring theory.

\paragraph{Acknowledgments}

 I thank  F.~Delduc  and B.~Vicedo for their comments. This review was written when visiting  the Albert-Einstein-Institut. I thank this Institute for its kind hospitality. This work is also partially supported by Agence Nationale de la Recherche under the contract ANR-07-CEXC-010.

\phantomsection
\addcontentsline{toc}{section}{\refname}

\end{document}